# Thinging Machine applied to Information Leakage

Sabah S. Al-Fedaghi
Computer Engineering Department
Kuwait University
Kuwait

Mahmoud BehBehani
Information Technology Department
Boubyan Bank
Kuwait

*Abstract*—This paper introduces a case study that involves data leakage in a bank applying the so-called Thinging Machine (TM) model. The aim is twofold: (1) Presenting a systematic conceptual framework for the leakage problem that provides a foundation for the description and design of a data leakage system. (2) The aim in (1) is developed in the context of experimentation with the TM as a new methodology in modeling. The TM model is based on slicing the domain of interest (a part of the world) to reveal data leakage. The bank case study concentrates on leakage during internal operations of the bank. The leakage spots are exposed through surveying data territory throughout the bank. All streams of information flow are identified, thus points of possible leakage can be traced with appropriate evidence. The modeling of flow may uncover possible hidden points of leakage and provide a base for a comprehensive information flow policy. We conclude that a TM based on the Heideggerian notion of thinging can serve as a foundation for early stages of software development and as an alternative approach to the dominant object-orientation paradigm.

I. INTRODUCTION

In software engineering, models have a central role in achieving a high level of reliability in the design, development, and deployment of systems. Specifically, in this context, we are interested in utilizing a conceptual model, the Thinging Machine (TM); reviewed in the next section), for the specification of the early phase of the life cycle of development in software systems. Without loss of generality, we focus on the problem of leakage of data. Accordingly, our aim is twofold:

*1)* Presenting a systematic conceptual framework for the leakage problem that provides a foundation for the description and design of a data leakage system. The framework is applied to an actual bank system.

*2)* The aim in (1) is developed in the context of experimentation with TM as a new methodology in modeling side by side with other methodologies such as object orientation. We will discuss the justification for pursuing such a venture later in this section.

A. *Background about Information Leakage*

Information or data (the terms are used interchangeably) leakage has a major impact on the business of many organizations today because valuable data are at risk of loss and possible exposure. The risks include loss of revenue and loss of credibility with customers, shareholders, or society. "Data leakage poses serious threats to organizations, including significant reputational damage and financial losses" [1]. The volume of data leakage has surpassed all expectations, mostly due to processing approaches where enterprises now centrally collect data, instead of keeping data details in various branches, thus maximizing big data benefits. [2].

According to [3], information leakage represents one of "the most common, but misunderstood, security risks faced by business and government alike." Firewalls, intrusion detection tools, and intrusion prevention mechanisms are deployed—"yet, the perception of the secure perimeter may be at odds with reality" [3]. "Despite a plethora of research efforts on safeguarding sensitive information from being leaked, it remains an active research problem" [1]. According to Lachniet [4], it is difficult to identify the requirements for data loss prevention, as well as "to whom and how they apply, and how to address them in a cost effective manner."

The 2017 Global Data Leakage Report [5], which is based on public information, includes the following data:

- The number of leaks increased by 37%.
- 60.5% of intruders were internal.
- 50.3% of violators were employees.

Information leakage is also a major concern to asset managers. In a recent survey, 35% of respondents claimed that information leakage represents the majority of their transaction costs. "Unfortunately, information leakage is hard to measure and harder to attribute to specific venues and behaviors" [6]. It is reported that 63% of the grayware (potentially unwanted programs that are not malicious and not viruses) applications in 2017 leaked phone numbers and 37% revealed the physical locations of phones [7].

Information leakage is a type of system vulnerability where sensitive data is released and such data can be useful for attackers to breach system security. A sample case of the problem is shown in the following scenario.

In many cases, the broker receiving the order is not the same broker who goes on to execute it on the chosen venue(s). As the originating broker, you give up control as soon as you pass an order on to another broker for execution. You do not necessarily know the route your client's order is taking, or how many other parties might get sight of that order—and how much information about that order leaks out—before it eventually hits the market. [8]

Another sample of information leakage involves printing devices. It was found that when a new printer was installed with large internal hard drives, accessible via IP, it retained information after printing jobs were completed, but it was not at all secure [3]. Leakage of information may not necessarily involve "loss" in the sense of *depriving* a victim from such a resource.

A third example is a 2009 case reported in the newspapers as follows. A member of parliament in Kuwait claimed he was in possession of hard evidence of financial irregularities by the prime minister. The lawmaker produced a $700,000 check signed by the prime minister favoring a former MP. The MP demanded to know the reason for handing the check to the lawmaker, insinuating possibilities of corruption. The prime minister's lawyer said he would file lawsuits against the MP for breaching bank confidentiality laws. The bank has also said it would file lawsuits against the MP and any employee who was involved in giving him a copy of the check.

Data leakage can happen because of internal and external breaches, either intentionally or inadvertently. It is reported that internal employees account for 43% of corporate data leakage, and half of these leaks are accidental [9]. "Accidental leaks mainly result from unintentional activities due to poor business process such as failure to apply appropriate preventative technologies and security policies, or employee oversight" [1].

Data leak prevention is the process of monitoring sensitive information, enforcing data handling policies, and assessing incidents of leakage. It is a strategy to ensure that such information does not reach the wrong hands during internal operations of an enterprise, either during communication outside it. The case study in this paper focuses on the former type of leakage.

Data leak prevention also refers to the use of technology products that assist in controlling the transferred data. According to Lachniet [4], "we must be concerned about controlling our sensitive data throughout its entire life-cycle (from creation to destruction)." Current approaches to data leak prevention systems are designed as risk reduction tools for specific hardware/software systems. Hardware/software platforms are installed on network links to analyze traffic for unauthorized transmissions, and they run on end-user servers that monitor data flow between users.

Many technical methods are used for leakage in enterprises [10]. For example, in relational databases, access behaviors are modelled in order to identify intrusions and detect data breaches [11]. Security data policies and traffic inspection can be utilized to protect sensitive information in communication and storage [12]. A typical tool used to handle data leakage is watermarking, where a unique code is implanted in the information container. Watermarks may require some alteration of the data and can sometimes destroy data. Moreover, the distributer (original owner) may have partners or may outsource where the data requires being shared [13].

### B. Aim and Approach with Regard to Information Leakage

We study the prospect that data has leaked along several points of the flow path, and we propose a flow-based model that facilitates the identification of leakages. The aim is to identify and monitor unintentional or deliberate disclosure of information in order to take appropriate steps to prevent any leak in enterprise environment.

In our proposed system, all streams of information flow are identified, thus pointing to potential leakages that can be traced with appropriate evidence. This modeling of flow may uncover possible hidden points of leakage and provide a base for a comprehensive information flow policy. For example, it can be used to draw the specification of the privileges of administrators and employees and the internal information flow among them.

### C. Aim with Regard to Exploring a New Modeling Methodology

As mentioned previously, presenting a systematic conceptual framework for the data leakage problem is developed in the context of experimentation with a TM as a new methodology in modeling, alongside other methodologies such as object orientation. Many researchers have extended the use of object-oriented software design languages such as UML in order to apply them at the conceptual level (e.g., [14]). Although the huge development efforts and time that have been invested in UML and object-orientation-based studies, tools and mechanisms are marvelous achievements, this ought not be considered as the final word and should not discourage new research such as TM that points in other directions or may enrich the object-oriented paradigm itself.

The TM model is a diagrammatic language that is founded on slicing the domain of interest (a part of the world) to "bring out" things so that we can perceive them (nearness [15]) through thinging (presencing [15]) and describes how these things behave. According to Malafouris [16], humans evolve by creating new things, which in turn transform the ways we sense the world. "This applies to the modern forager of digital information as it applies to the Paleolithic hunter-gatherer and tool-maker" [16].

The notion of a thing and thinging in general plays an important role in modeling, contending with the salience of the widely acclaimed significance of the word *object*, the term currently in vogue among most software engineers. Heidegger [15] analyzed what makes a thing different from an object; a thing is self-sustained, self-supporting, or independent—something that stands on its own. The condition of being self-supporting transpires by means of *producing* the thing. On the other hand, objects are things locked into their final forms, closed in upon themselves: "It is as though they had turned their backs on us" [17]. Ingold [17] described the difference:

> Using a square of paper, matchstick bamboo, ribbon, tape, glue and twine, it is easy to make a kite. Indoors, we were assembling an object. [In] a field outside, they suddenly leaped into action, twirling, spinning, nose-diving, and—just occasionally—flying. The kite that had lain lifeless on the table indoors had become a kite-in-the-air. It was no longer an object, if indeed it ever was, but a thing. As the thing exists in its thinging, so the kite-in-the-air exists in its flying. [17]

TM takes thinging as a basic conceptualization notion. TM modeling consists of an arrangement of machines, wherein each thing has its unique stream of flow. TM modeling puts together all of the things/machines required to assemble a system (a grand machine). Accordingly, an additional aim of this paper is to explore the TM model capabilities in developing the notion of information leakage.

In the next section, we present a review of the TM model (also called the Flow thing model) as it is introduced in several publications [18-24]. The example in the section is a new contribution. Section 3 focuses on our case study of a *bank as a thing*. Applying TM to data leakage is the topic of section 4.

## II. THINGING MACHINE

According to Richard [25], diagramming is a thinking tool that transforms abstract issues into intelligible and actionable forms. TM modelling utilizes an abstract thinging machine (hereafter, machine) with five stages of thinging as shown diagrammatically in Fig. 1. A thing things; that is, a thing creates, processes, receives, releases, and transfers things. A machine that handles things is itself a thing that is handled by other machines, as illustrated in Fig. 2 (left). Fig. 2 (right) shows the snake as a machine that processes a frog and simultaneously as a thing that flows to an owl. The TM model is a grand thing/machine that forms the thinging of a system. Thinging here refers to the creation, processing, receiving, releasing, and/or transferring of the system (grand machine) or any of its submachines.

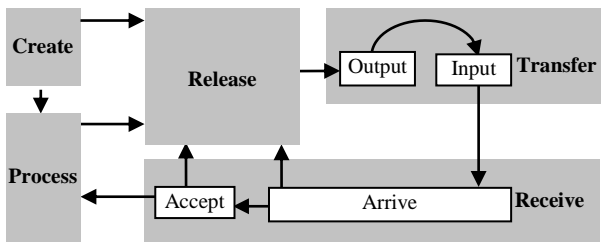

Fig. 1. Thinging Machine

Accordingly, a thing is a machine that manifests itself in the stages of creation, processing, receiving, releasing, and/or transferring, as shown in Fig. 1. The stages in the machine can be briefly described as follows.

**Arrive:** A thing flows to a new machine (e.g., packets arrive at a buffer in a router).

**Accept**: A thing enters a flow machine; for simplification purposes, we assume that all arriving things are accepted; hence, we can combine arrive and accept as the **receiving** stage.

**Release**: A thing is marked as ready to be transferred outside the machine (e.g., in an airport, passengers wait to board after passport clearance).

**Process** (change): A thing changes its form, but not its "identity" (e.g., a number changes from binary to hexadecimal).

**Create**: A new thing is born in a machine (e.g., a logic deduction system deduces a conclusion).

**Transfer**: A thing is input or output in/out of a machine.

TM includes one additional notation—triggering (denoted by dashed arrow)—that initiates a flow from one machine to another.

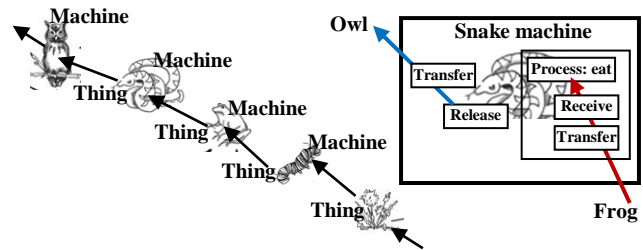

Fig. 2. Illustration of things that are machines and vice versa

**Example**: In contrast to object modeling, a *state* in TM is a thing with its own machine. Consider the classical object-oriented modeling of a coffee mug [26]. In object-oriented modeling, a coffee mug is an object with two states: *empty* and *filled*. TM takes a less abstracted view and considers a state as a submachine of a machine, as shown in Fig. 3. A mug (circle 1 in the figure) is a machine that involves the flow of coffee (2) that triggers the creation of the two states (3 and 4).

In such a scenario, we can identify four mutually exclusive *events*, as shown in Fig. 4. An event is a machine that is defined in terms of a time submachine and a region submachine (in addition to other machines). This notion of time as a thing/machine is not far from the Platonic view of time as a moveable image of eternity. Accordingly, the relevant events in the example are as follows:

Event a ($E_a$): Coffee is poured into the mug.
Event b ($E_b$): The mug is filled.
Event c ($E_c$): Coffee is poured out of the mug.
Event d ($E_d$): The mug is empty.

Accordingly, the behavior of the coffee/mug system is described as shown in Fig. 4. Any of the four events can be taken as the initial event. In the figure, time flow (transfer → receive → process [takes its course] → release → transfer) is not shown.

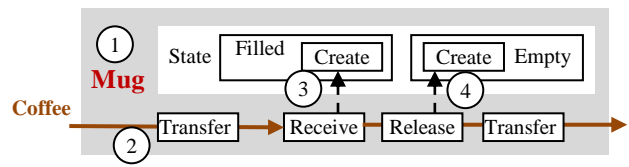

Fig. 3. The diagram of the system that involves filled and empty mug

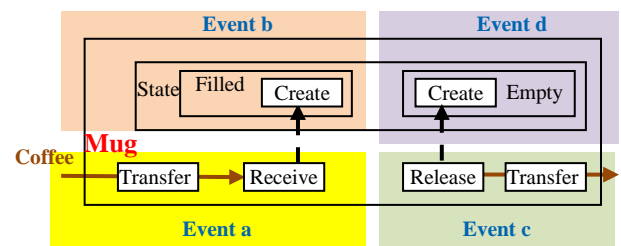

Fig. 4. Events

## III. BANK AS A THING

In this section, our study focuses on a bank as a thing. The bank banks through customers, other financial institutions, government agencies, and so forth. As these bank things encounter one another, the involved flows of different things/machines are interwoven and bundled together in a meshwork. The bank, in this picture, is a certain gathering together of the threads of the business world. It is modeled as a machine with many streams of flows that gather together the threads of banking.

In such a view, the bank switches from the usual perception of being an object to a "liveable" [17] thing that creates, processes, receives, releases, and/or transfers things. Remember, to objectify is to break a thing down into increasingly smaller parts instead of taking it holistically as it is. It is the "lifeless" kite object, as mentioned previously, that is viewed as a square of paper, matchstick bamboo, ribbon, tape, glue, and twine. Indoors, we were assembling an object. Additionally, the bank thing components do the same as exemplified by data leakage, the focus of this paper. In reality, any current bank as an object is a "livable" system to a certain degree, but this, by necessity, is an implicit result of its functions, and it is a partial "livability".

For example, leakage in the bank thing is a gathering of (sub)things and is viewed as one of the inhabitants of the bank, analogous to an octopus in an ecosystem with long arms that extend everywhere: employees, computers, desks, and cabinets, and so forth. This octopus is hiding until it is "brought out" by the bank's thinging. It can cause harm if not dealt with holistically. The main result of our case study in the next section is exposure of this leakage thing through mapping its territories (octopus arms) throughout the bank.

Such a perspective uncovers many hidden things as bank's dwellers. Our task is identifying these hidden occupiers of the bank in its model (diagram), as will be demonstrated by recognizing the information leakage thing. This is analogous to Wittgenstein's [27] work about the differences between "seeing" and "interpreting" (e.g., Wittgenstein's duck-rabbit figure), where in our study, assuming that our focus is on information leakage, we develop the bank's TM model, then we cut off the leakage machines inside it. This exposition of internally hidden machines is used for such purposes as constructing preventive measures and conducting forensics.

The bank as a thing in a modern society encounters ever-present vulnerability to threats. As a real thing, it is "a complicated machine in which every day something breaks down" [28]. A real bank is a gathering place that continuously calls for an unremitting effort to shore it up in the face of the comings and goings of its human inhabitants and nonhuman residents, not to mention the focus on security matters [17]. Much has been tried through developing a bank system that matches the expectations of well-ordered things within its outer boundaries; nevertheless, its function depends on the continual flow of things across these boundaries. The thingness of a bank becomes visible when an interruption or malfunctioning related to these flows appears.

In this paper, we focus on specific control efforts to counter the act of making information available without authorization. According to Lachniet [4], "Many controls are best done internally, such as creating a formal IT security management framework, or identifying the type of data you need to protect."

As used in this paper, leakage includes spilling, which refers to the unintended disclosure of information to unauthorized environments, organizations, or people [29]. In our study, we will exclude the situation of misconfigured systems that permit access to unprotected resources or are made available by hackers.

From the TM perspective, a leakage thing (e.g., information leakage) is a flow that spills out of the grand TM machine. This implies that a submachine has malfunctioned in the bank. Fig. 5 shows four possible types of submachine that leak flow from (1) received, (2) processed, (3) created, and (4) released information.

In the next section, we will identify all possible malfunctioned submachines in the bank used in our case study after developing a TM description of certain operations (e.g., consumer e-purchases) in the bank.

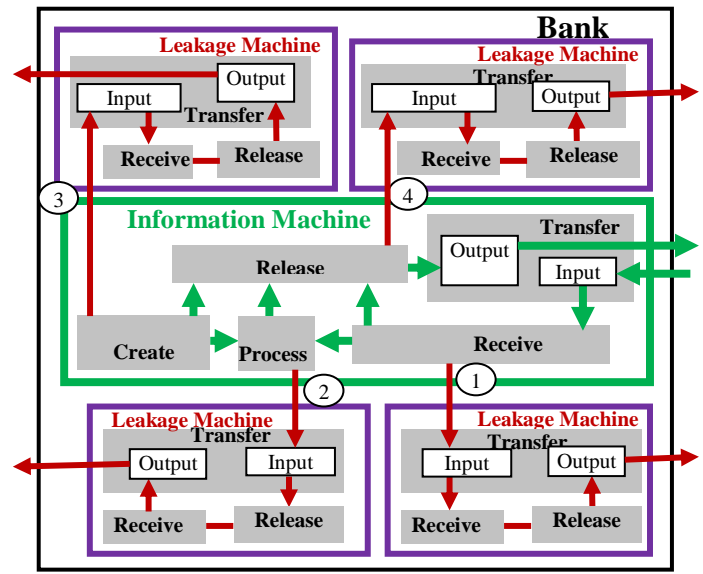

Fig. 5. Leak flows to leakage machines

## IV. BANK AS A CASE STUDY: CONSUMER E-PURCHASES

In the existing system of the bank in our case study (and in all banks, to the best of our knowledge), no explicit focus is given to the issue of information leakage. Such an issue is discreetly handled by the security team that deals with matters such as detecting hacking, collecting evidence, and the use of security tools such as encryption. In such a context, in specific leakage cases a possibility exists that the inability to progress legally due to a lack of a predesigned amount of evidence means the chances of escape for leakers are great.

To demonstrate the application of FM modeling in the area of a leakage of data, we took the following steps:

- First, we developed complete static and dynamic descriptions of the bank TM by focusing on the sample

application of e-purchases because of the paper size limitation.

- Then, we exposed leakage machines inside the bank TM description.
- 

A. *Consumer E-purchase*

As shown in Fig. 6, a customer (upper left corner) clicks on an icon on the screen that creates a signal (1) that flows to the electronic device software system to be processed (2) and triggers the processing (e.g., filling with relevant data) of a purchase request (3). Note that the request data is provided by the customer selection (click), and the (blank) purchase request is already stored in the device.

The request flows to the merchant server to be validated (4), and this triggers the generation of a formatted message (5) that flows to the payment gateway server (6). There, it is processed (7) to trigger the release of a processed payment page (8) according to the given data of the request. The payment page flows to the merchant server (9), then flows to the customer's electronic device where it is displayed (10). The customer inputs the payment details to trigger the creation of a transaction (11), which includes the card data, PIN number, and bank ID.

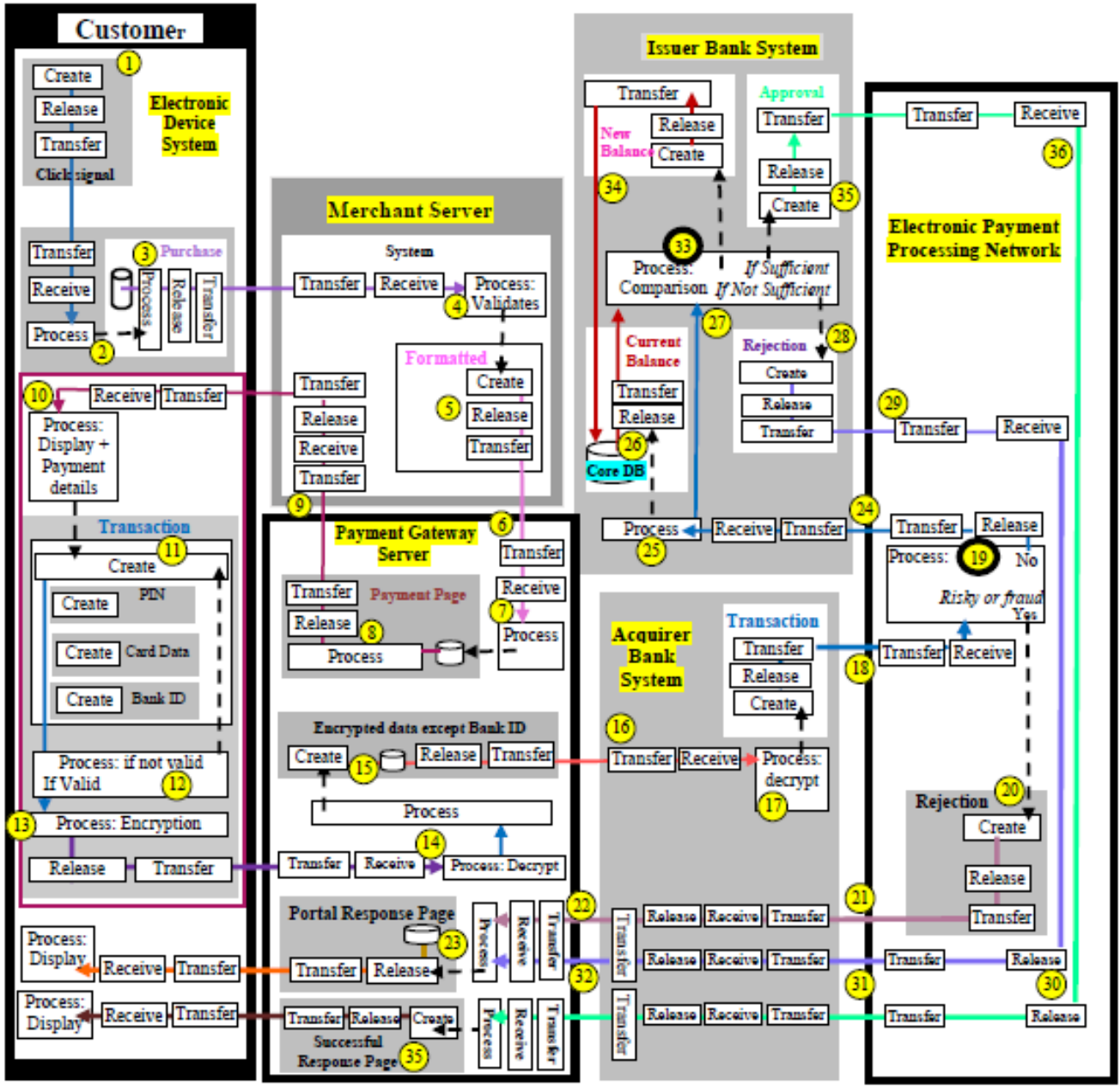

Fig. 6. The static description of a consumer e-purchase

To give a further description of how to create the transaction machine, Fig. 7 shows a sample of constructing it from clicks on the screen, a physical bank card, and a stored value in the inputting device. This thing-oriented depiction contrasts with the typical object-oriented specification (e.g., in UML), which has the mere structure of a class and its attributes.

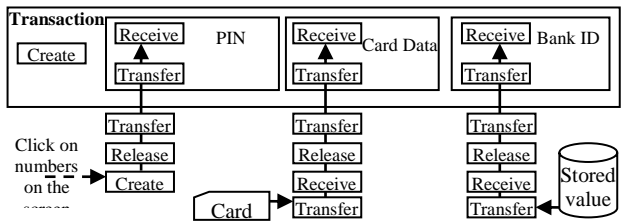

Fig. 7. Sample description of input

Continuing with Fig. 6, the transaction (11) is validated (processed) in the device browser such that

- If the input data is invalid (12), then the transaction is constructed again, or
- If the data is valid, then the transaction is encrypted (13) and flows to the payment gateway server (14).

In the payment gateway server, the encrypted transaction is received and decrypted (14). Then, the original transaction is processed to generate encrypted data without encrypting the bank ID (15). The encrypted data is stored into a database, and then released to the acquirer bank (16).

In the acquirer bank, the transaction data is processed into decrypted data (17). Then, the transaction flows to the payment processing network (18) where it is processed (19) and the following can happen:

- If the transaction is risky or possibly fraudulent, then a rejection response is generated (20) and sent to the acquirer bank (21) and then to the payment gateway server to be processed (22) to release a pre-stored portal page response from the database (23). The portal page response is transferred to the customer's device browser to be displayed.
- Returning back to the processing at (19), if the transaction in the electronic payment network is not considered a risk or possible fraud, then the transaction flows to the issuer bank (24) to be processed (25). This processing triggers a release of the customer's current balance from the centralized banking database (26) to be compared with the transaction amount that is received by the acquirer bank (27).
- If the balance is not sufficient, then a rejection response is created (28) and transferred to the electronic payment processing network (29), and then it flows to the acquirer bank (30), to the payment gateway server (31), and to the payment gateway server (32) to trigger a release of a pre-stored portal page response that is transferred to the customer's device browser.
- Going back to the comparison of the balance with the transaction cost, if the balance is sufficient, then a new balance is calculated (33) and stored in the core banking database (34). Additionally, an approval response is created (35) and flows across servers to create a success page response (36) as described before.

Fig. 6 gives a static description of a consumer e-purchase. To describe its dynamic behavior, we give the following events, as illustrated in Fig. 8, which is a copy of Fig. 6 marked with regions of events.

**Event 1 ($E_1$)**: The customer clicks on his/her electronic device browser, which is processed by the device.
**Event 2 ($E_2$)**: A purchase request is sent to the merchant where it is validated.
**Event 3 ($E_3$)**: The merchant sends a formatted message to the payment gateway server to be processed.
**Event 4 ($E_4$)**: The payment gateway server processes a stored page and sends it to the customer's browser through the merchant server.
**Event 5 ($E_5$)**: The portal payment page instructs the customer to insert his/her payment details.
**Event 6 ($E_6$)**: The payment details are inputted and validated in the electronic device browser.
**Event 7 ($E_7$)**: In case it is invalid, the browser requests that the customer re-input the correct payment information.
**Event 8 ($E_8$)**: The electronic device browser encrypts the transaction.
**Event 9 ($E_9$)**: The electronic device browser sends the transaction to the payment gateway server where it is decrypted.
**Event 10 ($E_{10}$)**: The payment gateway server processes the decrypted transaction by separating the bank ID.
**Event 11 ($E_{11}$)**: The payment gateway server encrypts the transaction, except for the bank ID, then stores it into the database.
**Event 12 ($E_{12}$)**: The encrypted data, with the exception of the bank ID, is transferred from the payment gateway server to the acquirer bank system, where it is decrypted.
**Event 13 ($E_{13}$)**: The decrypted transaction is generated in the acquirer bank and transferred to the electronic payment processing network, where is processed for possible fraud.
**Event 14 ($E_{14}$)**: If the transaction is fraudulent, then a rejection response is sent to the acquirer bank.
**Event 15 ($E_{15}$)**: The payment gateway server received the message from the acquirer bank and processes it.
**Event 16 ($E_{16}$)**: The electronic device browser displays the rejection message.
**Event 17 ($E_{17}$)**: The electronic payment processing network sends the transaction to the issuer bank to be processed.
**Event 18 ($E_{18}$)**: The issuer bank gets the customer's current balance and processes it against the transaction received.
**Event 19 ($E_{19}$)**: If the current balance is not sufficient, then the issuer bank generates a rejection response that flows to the electronic payment processing network then to the acquirer bank.
**Event 20 ($E_{20}$)**: If the current balance is sufficient, then the issuer bank deducts the requested amount from the customer balance.
**Event 21 ($E_{21}$)**: The issuer bank generates an approval message that flows to the electronic payment processing network then to the acquirer bank.
**Event 22 ($E_{22}$)**: The electronic device browser displays the portal payment success page.

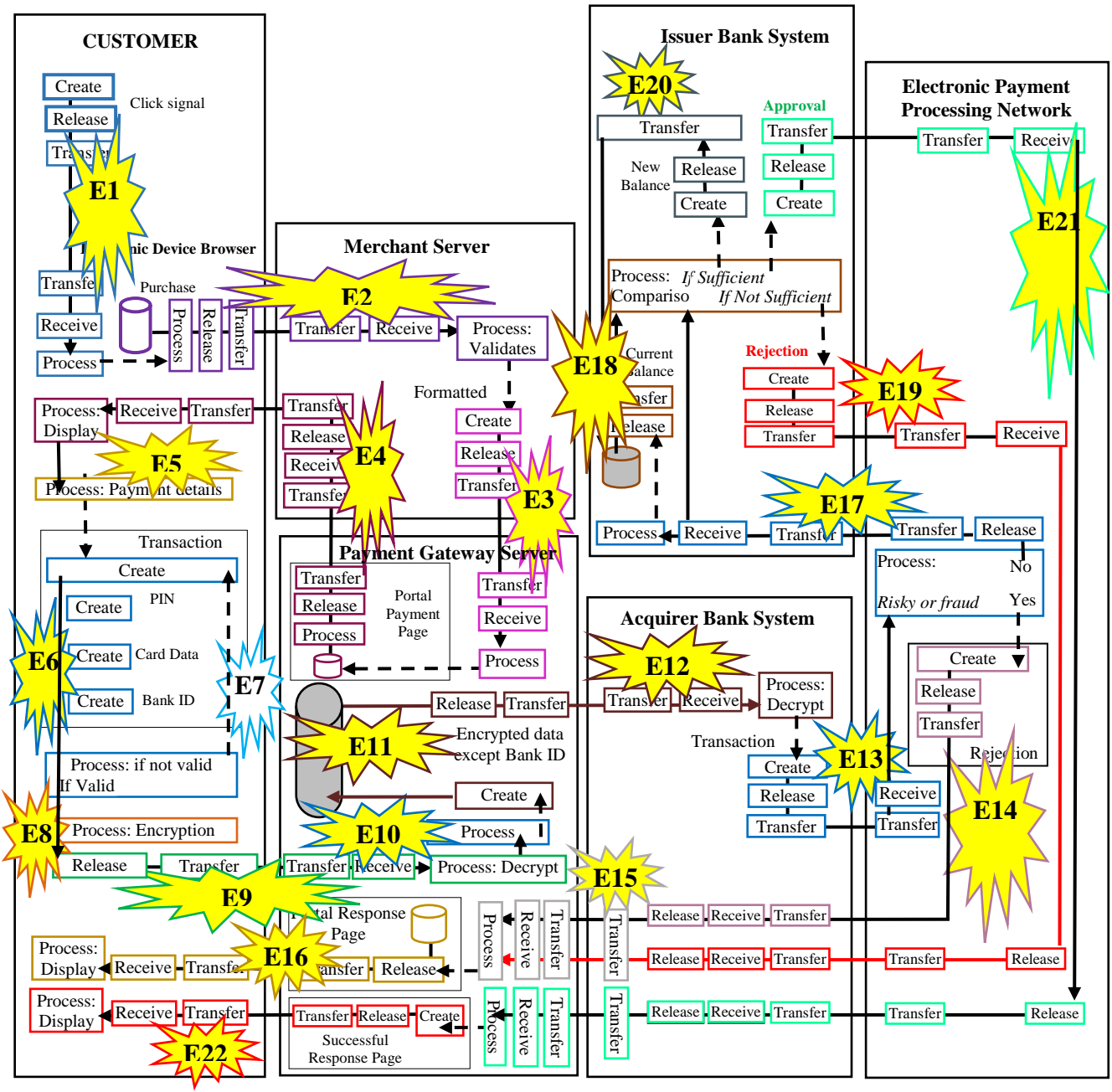

Fig. 8. The events of the consumer e-purchase

## V. Leakage Submachines

In this section, we will identify all possible leakage submachines in the bank. As an example of such identification, we focus on the issuer bank, keeping the numbered circles of Fig. 6. Fig. 9 shows the selected area for analyzing a data leakage. The aim is to model the entire issuing bank as a physical environment of the information system that handles the transaction data during its life cycle. The analysis can be generalized to different areas of consumer e-purchases.

Accordingly, Fig. 10 shows this expanded representation of the issuing bank. The top part of Fig. 10 shows the switch server where the transaction data flows from the electronic payment network to the issuer bank (24) to be processed (25). This processing triggers a release of a customer's current balance from the core database (26) to be compared with the transaction amount that is received by the acquirer bank (27).

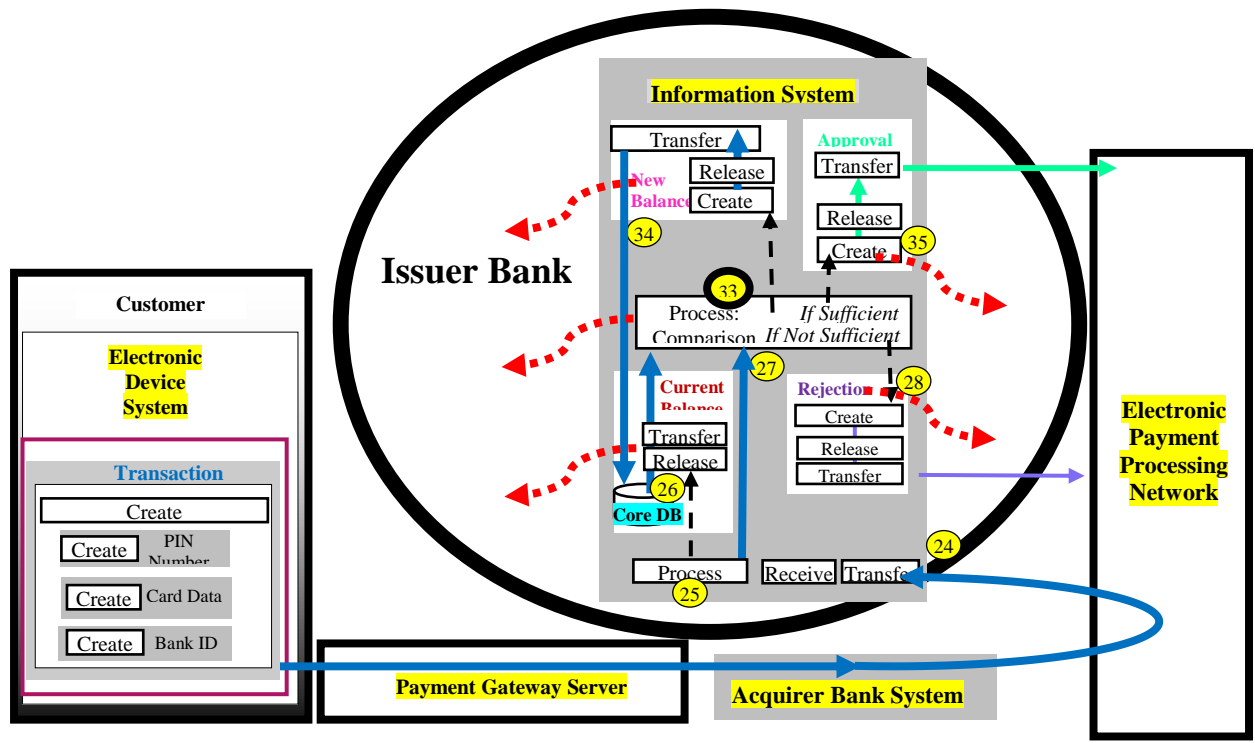

Fig. 9. The selected area for analyzing leakage

In addition to this main flow that involves the switch server and the core database, the server room includes the console, the core server, and the local e-mail server.

In Fig. 10, circles A and B in the server room point to possible leakage because the console screen can be captured by an employee (e.g., cleaning staff member).

The IT employee (C) monitors the system activity through his/her PC. He/she can access data on the switch server by creating an access request that flows to the core server. The currently processed transaction is copied and sent as an attachment (F) of an e-mail to the IT employee.

In the employee's PC (G), it is displayed, flows to the email server (F) and then printed (I). Here, there is an opportunity for leakage (e.g., by taking a picture or using flash drive). The hard copy is sent via a messenger employee to an employee who files it. Accordingly, we can describe all possible leakage machines, identifying their locations and who activates them. An employee can walk out of the bank carrying the data on a flash drive or as a hard copy. He/she can use the regular bank mail to send it out. These examples illustrate the method of identifying all possible leakage machines.

To summarize, Fig. 11 shows a general picture of different flows. The blue arrows in the figure show legitimate flows of the data whereas the red ones indicate leakage. The red flows originate from a leakage machines as follows.

Leakage machine 1 (circle 1): Capturing the console screen by an employee who has access to the server room.

Leakage machine 2 (circle 2): Capturing data from a PC by an IT employee using a camera, flash drive, etc.

Leakage machine 3 (circle 3): Copying data (hard copy) by an IT employee.

Leakage machine 4 (circle 4): Capturing data from a PC by a non-IT employee using a camera, flash drive, etc.

Leakage machine 5 (circle 5): Copying data by a record-keeping employee.

Leakage machine 6 (circle 6): Copying data by a messenger.

Leakage machine 7 (circle 7): An employee obtains a hard copy in an unauthorized way.

This thinging approach means that the leakage machine stands apart from its bank grand machine and is treated as a unified whole. A machine of interest (e.g. leakage) is exposed out of the bank thing with further thinging. It would appear as a subdiagram, in the forefront, clearly contrasted against the ground. This thinging of leakage is an act of "creation" of a machine which is already "exits" in reality even though we only perceive it when it becomes alive.

Such a comprehensive picture of data leakage provides the basis for planners and security personnel to focus on aspects that are suitable for the required prevention level. Additionally, it furnishes a foundation for any forensic investigation.

VI. CONCLUSION

This paper sought to accomplish two aims: present a systematic conceptual framework for the leakage and to develop that in the context of experimentation with TM as a new methodology in modeling. The TM model of the bank demonstrates the viability of the TM model.

The TM diagrams may look complex; however, they can be simplified by lumping the details together or omitting stages according to requirements. Many issues remain to be clarified; however, this paper demonstrates the potential feasibility of this approach.

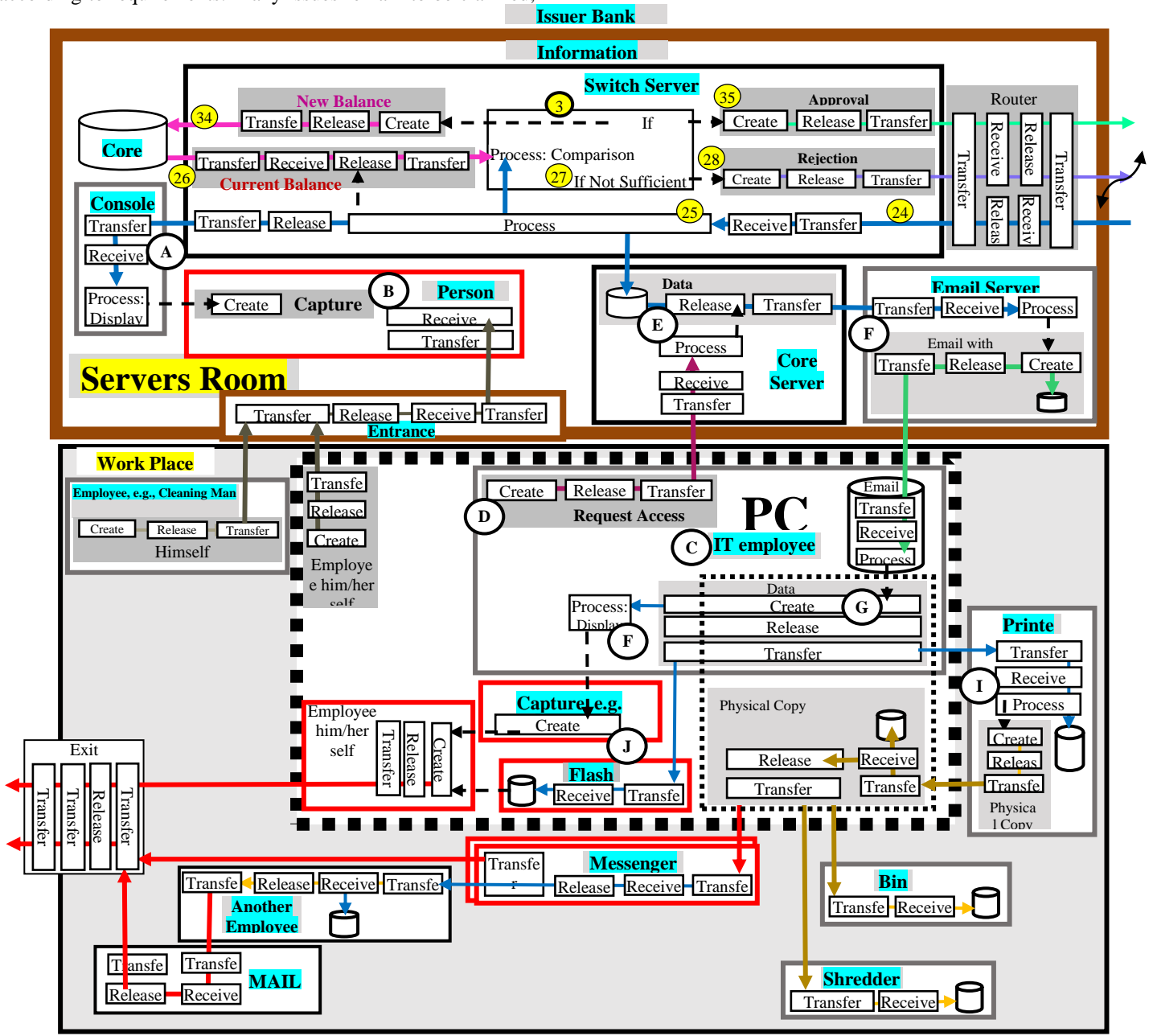

Fig. 10. Description of the flow of data and its physical environment

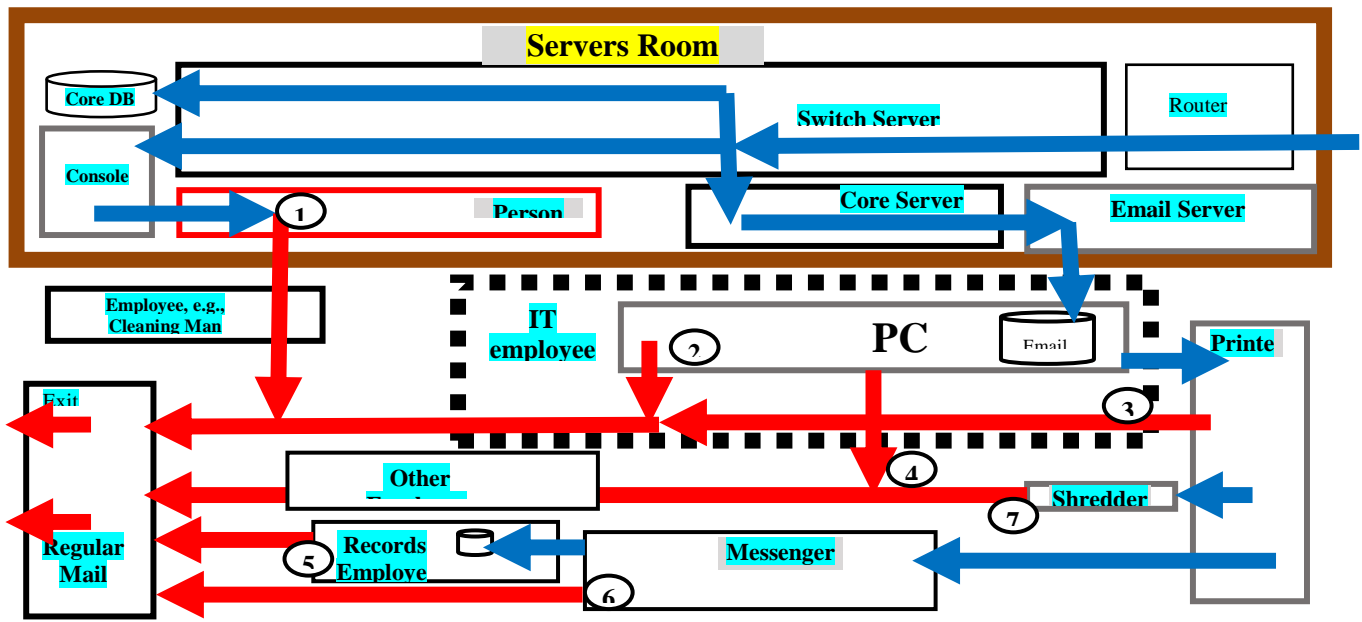

Fig. 11. A general picture of possible leakage in the example